\documentclass[journal]{IEEEtran}
\usepackage{blindtext}
\usepackage{subcaption}
\usepackage{graphicx}
\usepackage{bm}
\usepackage{amsmath}
\usepackage[colorlinks=false]{hyperref}

\begin{document}
%
\title{Fully-Tensorial Modeling of Stimulated Brillouin Scattering in Photonic Waveguides}

\author{Marcin Malinowski, and Sasan Fathpour, {\it Senior Member} 
\thanks{M. Malinowski is with CREOL, The College of Optics and Photonics, University of Central Florida, Orlando, Florida 32816, USA e-mail: marcinmalinowski@knights.ucf.edu}
\thanks{S. Fathpour is with CREOL, The College of Optics and Photonics, and Department of Electrical and Computer Engineering, University of Central Florida, Orlando, Florida 32816, USA}
\thanks{Manuscript received XXX, revised XXX}}%

\markboth{IEEE JOURNAL OF QUANTUM ELECTRONICS, Vol XXX, DATE XXX}%
{Shell \MakeLowercase{\textit{et al.}}: Bare Demo of IEEEtran.cls for Journals}

\maketitle

\begin{abstract}
A thorough study of elastic waves in waveguides, taking into account the full tensorial nature of the stiffness tensor, is presented in the context of stimulated Brillouin scattering. Various approximations of the elastic wave equation used in the stimulated Brillouin scattering literature are implemented and their validity and applicability are discussed. The developed elastic wave mode-solver is also coupled with an electromagnetic counterpart to study the influence of elastic anisotropies on Brillouin gain.
\end{abstract}


%
\IEEEpeerreviewmaketitle

\section{Introduction}
Stimulated Brillouin scattering (SBS) has been extensively studied in optical fibers, albeit it is generally considered a nuisance for the long-haul optical communication application \cite{ippen1972stimulated}. Recently, there has been renewed interest in SBS in order to harness it in integrated photonic devices \cite{eggleton2013inducing,kittlaus2016large, van2015interaction, choudhary2017high}. The narrow bandwidth of the SBS response leads to a host of applications in microwave photonics \cite{pant2014chip}, such as the construction of tunable bandpass \cite{byrnes2012photonic} and notch \cite{marpaung2015low} filters, phase shifters \cite{pagani2014tunable} and microwave synthesizers \cite{li2013microwave}. Also, long phonon lifetime has been used to store optical pulses in the acoustic domain \cite{zhu2007stored}. SBS has been, furthermore, utilized in wavelength-selective amplifiers \cite{olsson1987characteristics} and narrow-linewidth lasers \cite{stokes1982all}. Cascaded SBS process has been proposed as a multiwavelength source for optical communications \cite{al2005widely}. 

Unlike silica, the common material in optical fibers, most of the materials used in integrated photonics are not isotropic. A prime example is silicon with a cubic lattice structure \cite{hopcroft2010young}, on which large Brillouin gain has been demonstrated \cite{kittlaus2016large}. To fully explore the whole parameter space of integrated acoustooptic devices, knowledge of all acoustic modes is necessary, especially given that the acoustic and optical modes can be tailored independently, while retaining coupling between the two classes of modes as in the case of a Brillouin laser on silicon membranes \cite{otterstrom2017silicon}.

The finite-element method (FEM) is a versatile modeling choice for implementing an elastic-wave mode-solver that takes into account the tensorial stiffness of materials. The method has been originally  developed in the 1960s and 70s to model the problems of structural mechanics \cite{zienkiewicz1977finite}, including the elastic-wave equation \cite{castaings2008finite}, sometimes also called the seismic wave equation. Unlike the finite-difference method, FEM can easily handle complex geometries. Among other modeling problems, FEM has also been used to find the acoustic modes of structures in the ultrasound and the GHz ranges.  Examples are the scalar pressure model in chalcogenide waveguides \cite{poulton2013acoustic}, the scalar elastic model for SBS in fibers \cite{ward2009finite}, isotropic beams \cite{wilcox2002dispersion}, and the full-viscoelastic response of waveguides \cite{bartoli2006modeling} at ultrasound frequencies. The mode-solvers fall under the general scheme of the semi-analytical finite element method (SAFE), which reduces the three-dimensional (3D) wave-propagation problems to 2D variants, by assuming periodicity in the direction of propagation \cite{castaings2008finite}. In this paper, we develop and implement the SAFE method for fully-tensorial elastic-wave equation, described in Section III, using an open-source element-solver, called \textit{FEniCS} \cite{logg2010dolfin}. 

In Section IV on isotropic materials, we perform a thorough review of various approximations of the elastic-wave equation in the SBS literature and discuss their validity in the context of integrated waveguides. As an example, a material system  consisting of  a chalcogenide glass, As$_2$S$_3$ embedded in thermal oxide is studied \cite{pant2011chip}. So far, the consensus for such isotropic materials, and following the optical fiber literature \cite{boyd2003nonlinear}, has been to assume that the displacement in waveguides follows the scalar acoustic-wave model \cite{poulton2013acoustic}. However, we find that this model overestimates the eigenmodes of the system by as much as 0.5 GHz in submicron-sized waveguides. In addition, the acoustic-wave equation assumes close to plane-wave propagation. A better approximation is provided in reference \cite{ward2009finite}, which we later refer to as the  scalar elastic equation. In both cases, nonetheless, these models do not appropriately capture the mode profile near the material interfaces, when compared with the most general model used in this work. 

In Section V on anisotropic materials, we couple the elastic mode-solver with an electromagnetic counterpart to explore the effect of elastic anisotropies on Brillouin gain, in addition to the variation of the mechanical resonance frequency.  Silicon is chosen as an example, which is ubiquitous in integrated photonics and provides high Brillouin gain in suspended structures \cite{kittlaus2016large,van2015interaction}. Here, researchers have resorted to full-3D simulations with the Floquet boundary conditions \cite{sarabalis2016guided, smith2016stimulated} usually implemented in the commercial FEM solver, COMSOL$^{TM}$. Following the micro-electro-mechanical systems (MEMS) literature \cite{hopcroft2010young}, some authors have chosen to use the simplified isotropic model for silicon \cite{rakich2012giant, qiu2013stimulated}. While qualitatively acceptable, this assumption can lead to large discrepancies quantitatively. We show that in extreme cases, where the waveguide is aligned along the [100] or the [110] crystalline axis, the difference in eigenfrequencies can be as high as 0.8 GHz, which highlights the need for a fully-tensorial formulation. This is directly translated to the Brillouin gain, which scales with the inverse of the mechanical resonance-frequency squared. For back-scattering, we show that, for an arbitrary crystal orientation in silicon, the elastic modes do not need to have the same symmetry as the optical modes, which greatly affects the Brillouin coupling coefficient. While, in principle, the anisotropic behavior of silicon can be simulated in COMSOL$^{TM}$, the use of the SAFE method in this work leads to substantial - over two orders of magnitude - computational-time improvement.   Furthermore, since the SAFE method reduces the equations to a 2D problem, it requires less computational memory than the full-3D simulation, without neglecting any physical modeling features.  Finally, it should be noted that recently a mode-solver for isotropic materials utilizing the finite-difference method has been presented \cite{dostart2017elastic}, but our work provides faster convergence through the use of higher-order finite elements.

\section{Governing Equations}
The constitutive equation of motion for an elastic medium is \cite{zienkiewicz1977finite} 
\begin{equation}
\nabla \cdot \boldsymbol {\sigma} = \rho\frac{\partial^2\boldsymbol{u}}{\partial^2 t},
\label{eq:conservation_of_momentum}
\end{equation} where $\boldsymbol{u}$ is the displacement vector, $\rho$ is the density and the  $\boldsymbol {\sigma}$ is the stress. Stress is linearly related to strain,  $\boldsymbol{\epsilon}$, via the stiffness tensor, $\boldsymbol{C}$, as follows:
\begin{equation}
 \boldsymbol {\sigma} = \boldsymbol{C} : \boldsymbol{\epsilon} \quad\text{and}\quad  \boldsymbol{\epsilon} = \frac{1}{2}[\nabla  \boldsymbol{u} + (\nabla  \boldsymbol{u})^T].
\end{equation}
To obtain the equations describing the modal distribution, the Fourier transforms of {\it t} and {\it z} are employed. We also introduce the transverse gradient operator $\nabla_T$ such that $\nabla = \nabla_T + iq_b \boldsymbol{\hat{z}},$
where $q_b$ is the acoustic-wave propagation constant. Afterwards, the elastic wave equation reads \cite{castaings2008finite, wolff2015stimulated}
\begin{equation}
\begin{split}
&\nabla_T \cdot \boldsymbol{C} : \frac{1}{2}[\nabla_T \boldsymbol{u} + (\nabla_T \boldsymbol{u})^T] + iq_b \boldsymbol{\hat{z}} \cdot \boldsymbol{C} : \frac{1}{2}[\nabla_T \boldsymbol{u} + (\nabla_T \boldsymbol{u})^T] + \\
&\nabla_T \cdot \boldsymbol{C} : \frac{1}{2}[iq_b \boldsymbol{u} + (iq_b \boldsymbol{u})^T] + iq_b \boldsymbol{\hat{z}} \cdot \boldsymbol{C} : \frac{1}{2}[iq_b \boldsymbol{u} + (iq_b \boldsymbol{u})^T]  = \\
& -\rho \omega^2 \boldsymbol{u}.
\end{split}
\label{eq:elastic_wave}
\end{equation}
To obtain the weak form of Equation \ref{eq:elastic_wave}, we multiply 
it by a test function $\boldsymbol{v^*}$, integrate and apply the Green's theorem in analogy to the 3D case \cite{johnson2012numerical}, to obtain
\begin{equation}
\begin{split}
&-\int_\Omega \boldsymbol{\epsilon}_T(\boldsymbol{v^*}):\boldsymbol{C}:\boldsymbol{\epsilon}_T(\boldsymbol{u})dx + \int_\Omega \boldsymbol{\epsilon}_Z(\boldsymbol{v^*}):\boldsymbol{C}:\boldsymbol{\epsilon}_T(\boldsymbol{u})dx  \\
+&\int_\Omega \boldsymbol{\epsilon}_T(\boldsymbol{v^*}):\boldsymbol{C}:\boldsymbol{\epsilon}_Z(\boldsymbol{u})dx + \int_\Omega \boldsymbol{\epsilon}_Z(\boldsymbol{v^*}):\boldsymbol{C}:\boldsymbol{\epsilon}_Z(\boldsymbol{u})dx  \\
&+\int_{\partial \Omega} \boldsymbol{v^*}\cdot[\boldsymbol{\hat{n}} \cdot \boldsymbol{C}:\boldsymbol{\epsilon}_T(\boldsymbol{u})]dx = 
-\int_\Omega \rho \omega^2 \boldsymbol{v^*} \cdot \boldsymbol{u} dx
\end{split},
\label{eq:weak_form}
\end{equation}
where the symmetry of the stiffness tensor $\boldsymbol{C}_{ijkl} = \boldsymbol{C}_{jikl}$, the notations $\boldsymbol{\epsilon}_T(\boldsymbol{u}) = \frac{1}{2}[\nabla_T \boldsymbol{u} + (\nabla_T \boldsymbol{u})^T]$, as well as $\boldsymbol{\epsilon}_Z(\boldsymbol{u}) = \frac{1}{2}[iq_b \boldsymbol{u} + (iq_b \boldsymbol{u})^T]$, have been utilized.  Also, $\Omega$ denotes the whole space,  $\partial \Omega$ the boundary and $\boldsymbol{\hat{n}}$ the normal to that boundary. Finding the elastic modes is equivalent to finding the eigenvectors of Equation \ref{eq:weak_form}, where $\omega^2$ is the eigenvalue.

\section{Implementation}

\subsection{Description}
\textit{FEniCS} is an open-source FEM solver that automates  significant portion of the finite-element assembly. The package contains the unified form language (UFL) \cite{alnaes2014unified}, a domain-specific language for declaration of variational forms with syntax that follows mathematical notation. The core of our code consists of Equation \ref{eq:weak_form}, implemented in UFL. The notable difference is that UFL does not support complex numbers, thus Equation \ref{eq:weak_form} is split into the real and imaginary parts that form a linear system of coupled equations. The Dolfin package \cite{logg2010dolfin} provides a high-level interface to various linear algebra packages that are need for efficient solution of the eigenvalue problem.  The FIAT package \cite{kirby2004algorithm} enables quick testing of different finite elements. For the tensorial elastic mode-solver, we have used the Lagrange finite elements and found them to be stable. 

When the eigenvectors are expressed in terms of the basis functions of the finite elements, $\boldsymbol{u} = u_i \phi_i$ in Equation \ref{eq:weak_form} becomes a generalized eigenvalue problem of the from $S(\phi_i, \phi_j)u_i = -M_{ij}\omega^2 u_i$, where the eigenvalues are weighted by the mass matrix, \textbf{M}, in our case being the material density. After splitting the complex coefficients into the real and imaginary parts, the weak form (Equation \ref{eq:weak_form}) forces the stiffness matrix, \textbf{S}, to be symmetric, which ensures that $\omega$ is real. The stiffness matrix is sparse, since $S(\phi_i, \phi_j)$ is zero for nonadjacent finite elements. While solving the eigenvalue problem, we are interested only in a few modes close to the fundamental mode, therefore there is no need to find all the eigenvectors. Given these requirements, the appropriate linear algebra package is recognized to be SLEPc \cite{hernandez2005slepc} with the Krylov-Shur algorithm and a spectral shift-and-invert preconditioner. SLEPc enables the parallel computation of eigenvalues, which is necessary for higher resolution. On a personal computer, with 16 GB of memory, we were able to compute eigenmodes for meshes up to 180 second-order Lagrange elements per facet. 

Both the stress and the strain tensors are symmetric, therefore to minimize memory usage we use the Voigt notation \cite{zienkiewicz1977finite}, where the two tensors are mapped onto six-dimensional vectors, i.e., the stress-strain relation becomes
\begin{gather}
 \begin{bmatrix} \sigma_{xx}  \\ \sigma_{yy} \\ \sigma_{zz} \\ \sigma_{yz} \\ \sigma_{xz} \\ \sigma_{xy}      \end{bmatrix}
 =
  \begin{bmatrix}
   c_{11} & c_{12} & c_{13} & c_{14} & c_{15} & c_{16} \\
   c_{21} & c_{22} & c_{23} & c_{24} & c_{25} & c_{26} \\
   c_{31} & c_{32} & c_{33} & c_{34} & c_{35} & c_{36} \\
   c_{41} & c_{42} & c_{43} & c_{44} & c_{45} & c_{46} \\
   c_{51} & c_{52} & c_{53} & c_{54} & c_{55} & c_{56} \\
   c_{61} & c_{62} & c_{63} & c_{64} & c_{65} & c_{66} \\
   \end{bmatrix}
    \begin{bmatrix} \epsilon_{xx}  \\ \epsilon_{yy} \\ \epsilon_{zz} \\ 2\epsilon_{yz} \\ 2\epsilon_{xz} \\ 2\epsilon_{xy}      \end{bmatrix}.
\end{gather}

Better numerical performance is obtained by using small numerical values, so dimensions of the parameters used in the solver are $[\rho]=$ g/cm$^3=10^{-12}$ g/$\mu$m$^3$, $[\boldsymbol{C}]=$ GPa $ = 10^{-12}$ g/(ns$^2 \mu$m) , $[x, \boldsymbol{u}] =\mu$m, which result in eigenvalues $[\omega]=$ GHz after cancellation of prefactors.  It is finally reminded that the mode-solver finds two modes for each eigenvalue, corresponding to waves traveling in opposite directions \cite{castaings2008finite}.

\subsection{Testing}
In the case of homogeneous isotropic materials, Equation \ref{eq:conservation_of_momentum} can be rewritten as $
 -\rho \omega^2 \boldsymbol{u} =   (2\mu + \lambda) \nabla(\nabla \cdot \boldsymbol{u}) -  \mu \nabla \times \nabla \times \boldsymbol{u} $. Conveniently, for shear waves $\nabla \cdot \boldsymbol{u} = 0$ and the problem reduces to the Maxwell's eigenvalue problem. If facets at $\pm a$ are clamped, then one of the solutions is
$u_y = \cos(k_x x)\exp(iq_bz)$, with $u_x,u_z = 0$ and for the given boundary conditions $k_x = 2\pi /a$. 
We use this fundamental mode to test the accuracy of the solver against a known solution. For the numerical simulations, we choose a 2$\mu$m $\times$ 5$\mu$m waveguide of thermal oxide and $q_b = 11.67 \mu$m$^{-1}$.

\begin{figure}[!h]
\centering
   \includegraphics[width=1\linewidth]{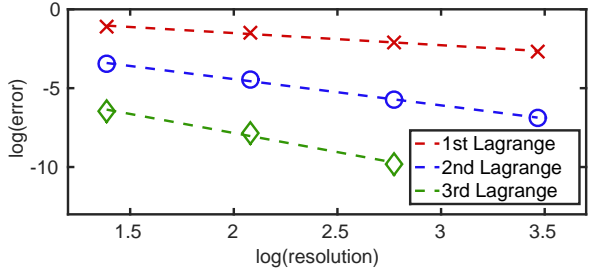}  
\caption{The error between the numerical and analytical eigenvalues plotted for Lagrange finite elements of different polynomial order. The convergence slopes for first, second, and third-order Lagrange elements are  -0.77, -1.66 and -2.42, respectively.}
   \label{fig:convergence}  
\end{figure} 

The error is estimated as $|\omega^2_{analytical} - \omega^2_{numerical}|$. This error is plotted against the resolution of the mesh for Lagrange elements of different polynomial orders \cite{logg2012automated} in Fig.
 \ref{fig:convergence}.  As expected, for third-order Lagrange elements, the convergence rate exceeds the second-order finite-difference scheme \cite{dostart2017elastic}.


\section{Isotropic materials:chalcogenides}
As the first example, the material parameters of a chalcogenide glass composition, As$_2$S$_3$, embedded in thermal oxide are implemented \cite{pant2011chip}. Both materials are isotropic, hence the non-zero components of their stiffness tensor elements are $c_{11} = c_{22} = c_{33} = 2\mu + \lambda$, $c_{44} = c_{55} = c_{66} = \mu$.  Also, the non-diagonal elements $c_{12}, c_{13}, c_{23} = \lambda$ together with their mirrored components. For thermal oxide, $\mu = 29.9$ GPa, $\lambda=$15.4 GPa and $\rho = 2.2$ g/cm$^{3}$ \cite{kim1996influence}, while As$_2$S$_3$ is expectedly softer with $\mu = 6.2$ GPa, $\lambda= 9.78$ GPa and $\rho = 3.2$ g/cm$^{3}$ \cite{pant2011chip, glaze1957properties}.



\begin{figure}[!h]
\centering
\vspace{-0.5cm}
\begin{subfigure}[b]{0.55\textwidth}
   \includegraphics[width=1\linewidth]{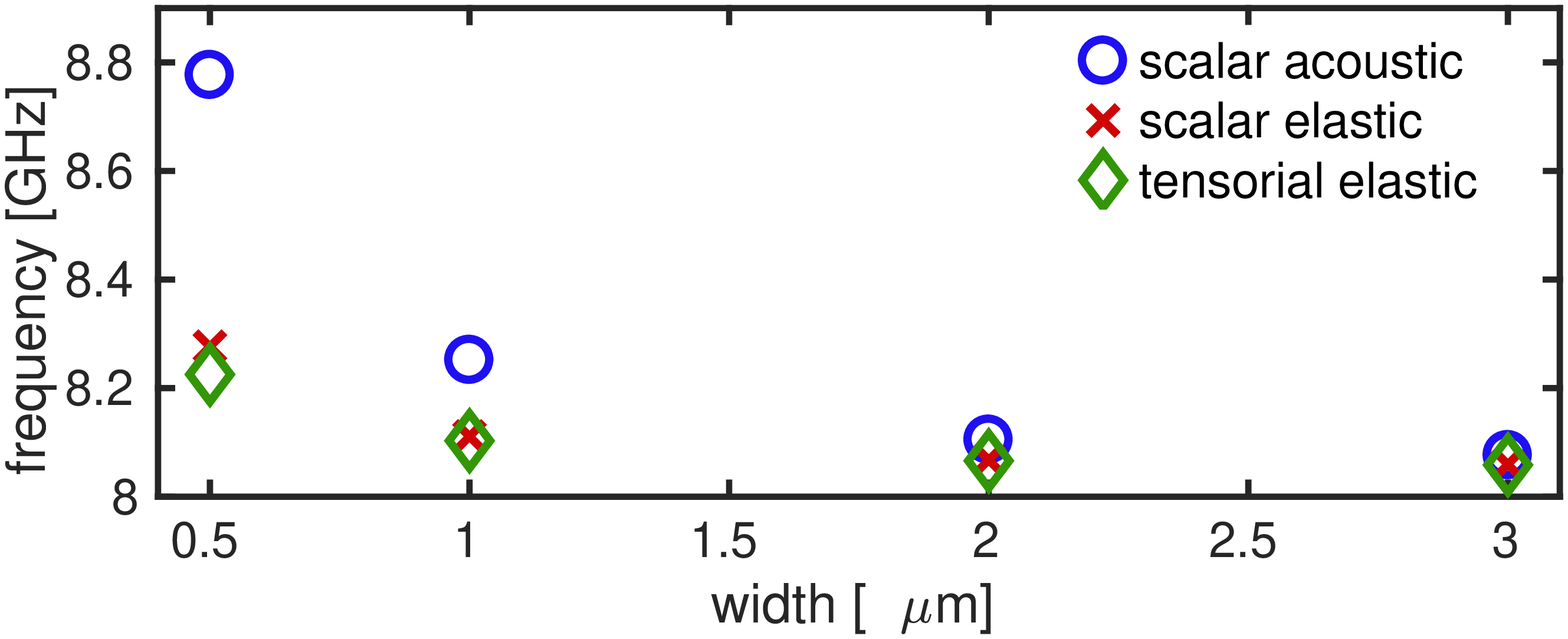}
   \caption{}
   \label{fig:equations} 
\end{subfigure}
\vspace{-0.5cm}
\begin{subfigure}[b]{0.55\textwidth}
  \includegraphics[width=1\linewidth]{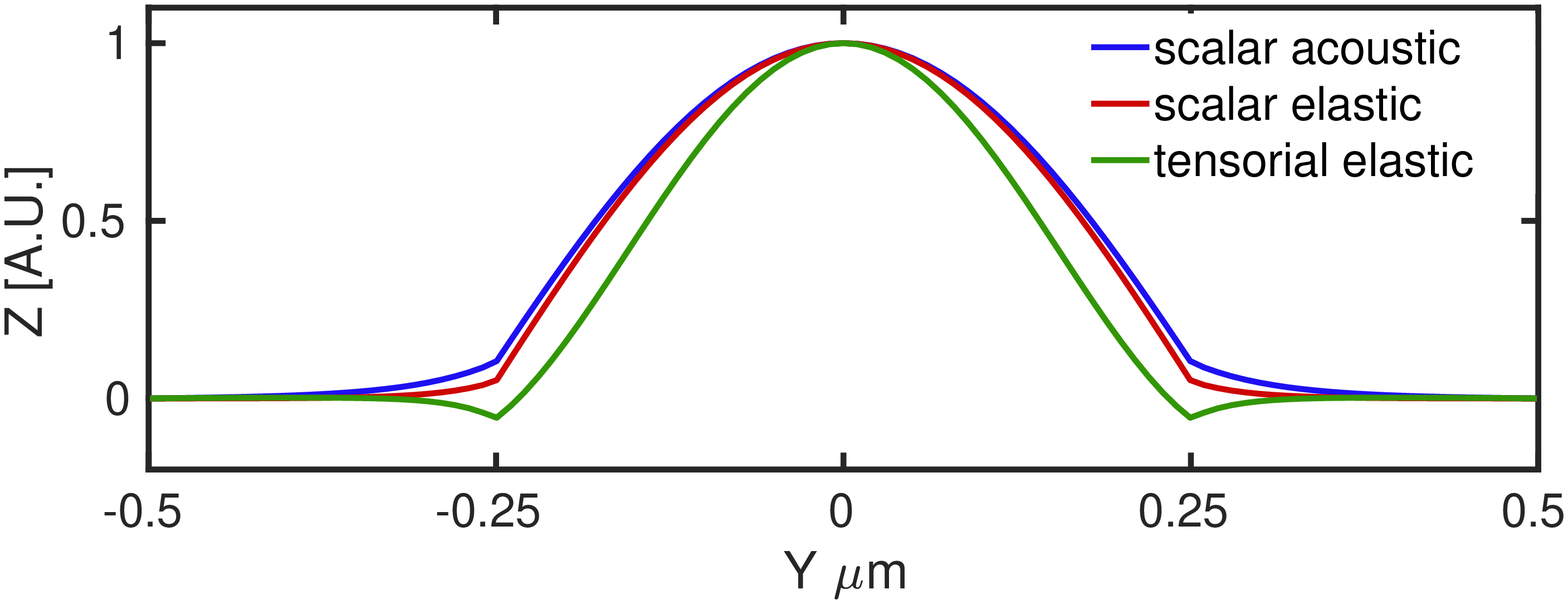}
   \caption{}
   \label{fig:cross_sections}
\end{subfigure}

\caption{ (a) The eigenvalue of the fundamental longitudinal mode of a square chalcogenide waveguide embedded in thermal oxide calculated using the scalar acoustic Equation \ref{eq:acoustic_model}, the scalar elastic Equation \ref{eq:scalar_elastic} and the full-tensorial elastic Equation \ref{eq:elastic_wave}. The computation was performed on the same mesh for each waveguide width. The acoustic model is too simplistic for waveguides with tight confinement. However, the scalar elastic equation remains accurate; (b) The cross-section of the eigenmodes calculated for 0.5-$\mu$m-wide waveguides using the three models. The plot shows the {\it Z}-displacement normalized to unity.}
\end{figure}

In the below comparison, the focus is on the fundamental longitudinal mode, $\boldsymbol{u} \approx u_z$. Following the optical fiber
literature \cite{boyd2003nonlinear}, the consensus so far has been to assume that the displacement in waveguides follows the scalar acoustic model \cite{poulton2013acoustic}
\begin{equation}
v_l^2 \nabla^2_T \tilde{\rho} - q_b^2 v_l^2 \tilde{\rho} = -\omega^2 \tilde{\rho},
\label{eq:acoustic_model}
\end{equation}
where the longitudinal velocity is related by $\rho v_l^2 = 2\mu + \lambda $ to the stiffness constants,
and $\tilde{\rho}$ refers to the change of material density. Since the displacement is only in the {\it z} direction, $\tilde{\rho}$ is proportional to $\rho u_z$, which implies that Equation \ref{eq:acoustic_model} can be rewritten as
$(2\mu + \lambda ) \nabla^2_T u_z - q_b^2 (2\mu + \lambda ) u_z = - \rho \omega^2 u_z$.
The eigenvalues of this and preceding equation are indeed identical. The issue with the acoustic model, however, is that it was originally developed for fluids and therefore neglects shear waves, or components of thereof. A better approach is to start with Equation 5 and
enforce $\boldsymbol{u} = u_z$, upon which we arrive at
\begin{equation}
\mu \nabla^2_T u_z - q_b^2 (2\mu + \lambda ) u_z = - \rho \omega^2 u_z,
\label{eq:scalar_elastic}
\end{equation}
which is known as the scalar elastic equation and has been previously used to model SBS in fibers \cite{ward2009finite}.  

Figure \ref{fig:equations} presents the fundamental longitudinal eigenvalues of chalcogenide waveguides, based on the three models (namely, Equations \ref{eq:elastic_wave}, \ref{eq:acoustic_model} and \ref{eq:scalar_elastic}) on the same mesh with second-order Lagrange elements. In all cases, the acoustic propagation constant is fixed to  $q_b = 18.23$ $\mu$m$^{-1}$, hence the eigenfrequency increases with mode confinement. The figure clearly shows that the scalar acoustic model is too simplistic for integrated devices with a discrepancy of 550 MHz for waveguides that are 0.5 $\mu$m wide. However, the scalar elastic equation remains accurate to within 100 MHz of the tensorial model for waveguides wider than 1 $\mu$m. The modal cross-sections normalized to unity are shown in Figure \ref{fig:cross_sections}, which shows a large discrepancy between the scalar models and the tensorial model near the material interfaces. 

	Based on the detailed model presented later in section V B, we also calculate the Brillouin gain for chalcogenide waveguides. The physical parameters used in the simulation are as follows: photoelastic constants of $p_{11}$ = 0.25 $p_{12}$	=  0.24 for As$_2$S$_3$,   and $p_{11}$ = 0.12 $p_{12}$ =  0.27  for SiO$_2$ \cite{galkiewiczAs2S3}, waveguide dimensions of 4 $\mu$m by 0.85 $\mu$m and experimental $Q_{mech}$ = 226 \cite{pant2011chip}. The phase-matching for back-scattering of the fundamental transverse-electric (TE) mode at 1544 nm dictates $q_b$ = 18.37 $\mu$m$^{-1}$.
Using the mode profile calculated using our solver, we obtain a gain coefficient of 322 W$^{-1}$m$^{-1}$, which agrees well with the scalar formulation, 321 W$^{-1}$m$^{-1}$, and the experimentally measured value of 311 W$^{-1}$m$^{-1}$ \cite{pant2011chip}. We note that for such large waveguides, the coupling is dominated by eletrostriction and the radiation pressure is negligible, that is less than 1 W$^{-1}$m$^{-1}$.

\begin{figure}[!h]
   \includegraphics[width=1\linewidth]{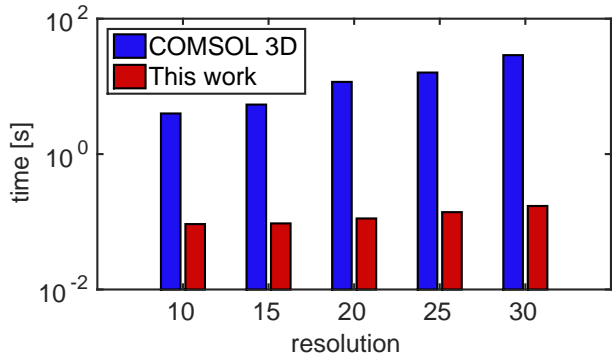}

\caption{A comparison of simulation time between the mode-solver in this work, and a fully-3D simulation in COMSOL$^{TM}$, confirming that the present solver is over 100 times faster. Additionally, the computational time increases faster with resolution for the 3D simulation than for the 2D SAFE method used in this work. The simulation was performed for chalcogenide waveguides embedded in oxide and backward scattering modes. In both cases, second-order Lagrange elements were used and 10 modes were calculated. COMSOL$^{TM}$ was called using the LiveLink$^{TM}$ Matlab interface for timing purposes. Resolution is understood as number of finite elements per facet. In the 3D simulation, Floquet boundary conditions were used and the size of elements in the {\it z}-direction was the same as in the {\it xy} plane. }
   \label{fig:speed}
\end{figure}

Finally, the chalcogenide glass material system is used to highlight the computational benefits of using the SAFE method against a fully-3D simulation in COMSOL$^{TM}$ with Floquet boundary conditions. Figure \ref{fig:speed} summarizes how this work provides two orders of magnitude decrease in computational time.

\section{Anisotropic materials:silicon}
\subsection{Resonance frequency}
Due to its high refractive index, it is anticipated that silicon should provide high Brillouin gain.  Until recently, observing the gain has been elusive, because the acoustic velocity in the silica layers, typically used for cladding the waveguides, is lower than the acoustic velocity of silicon.  In other words, the acoustic is leaky in standard silicon-on-insulator (SOI) waveguides. The solution to this problem is to use a suspended membrane \cite{kittlaus2016large} or a waveguide supported by a nanoscale strut \cite{van2015interaction}. It should be stressed that, due to the enhancement coming from radiation pressure, the strongest Brillouin gain in silicon arises from forward-propagating modes \cite{rakich2012giant}. 

Herein, the solutions of our solver are compared against the 3D finite-element solution of COMSOL$^{TM}$.  A rectangular silicon waveguide (450 nm $\times$ 230 nm) suspended in air is utilized, which means that $\int_{\partial \Omega} dx = 0$ in Equation \ref{eq:weak_form}. Also, the phase matching condition requires that the elastic wavevector be equal to the difference between the optical wavevectors, but to a good approximation the phonons oscillate in the {\it xy}-plane, thus we can look for modes with $q_b = 0$. Due to the symmetries of the stiffness tensor of silicon, a cubic material, the following identities hold true: $c_{11} = c_{22} = c_{33} = 164$ GPa, $c_{44} = c_{55} = c_{66} =  79$ GPa and the non-diagonal $c_{12}=  c_{13} =  c_{23} = c_{21} = c_{31} = c_{32} = 64$ GPa, while the rest of elements of the stiffness tensor are zero \cite{van2015interaction}. The density of silicon is 2.328 g/cm$^3$.  

\begin{figure}
\centering
\vspace{-0.5cm}
\begin{subfigure}[b]{0.55\textwidth}
   \includegraphics[width=1\linewidth]{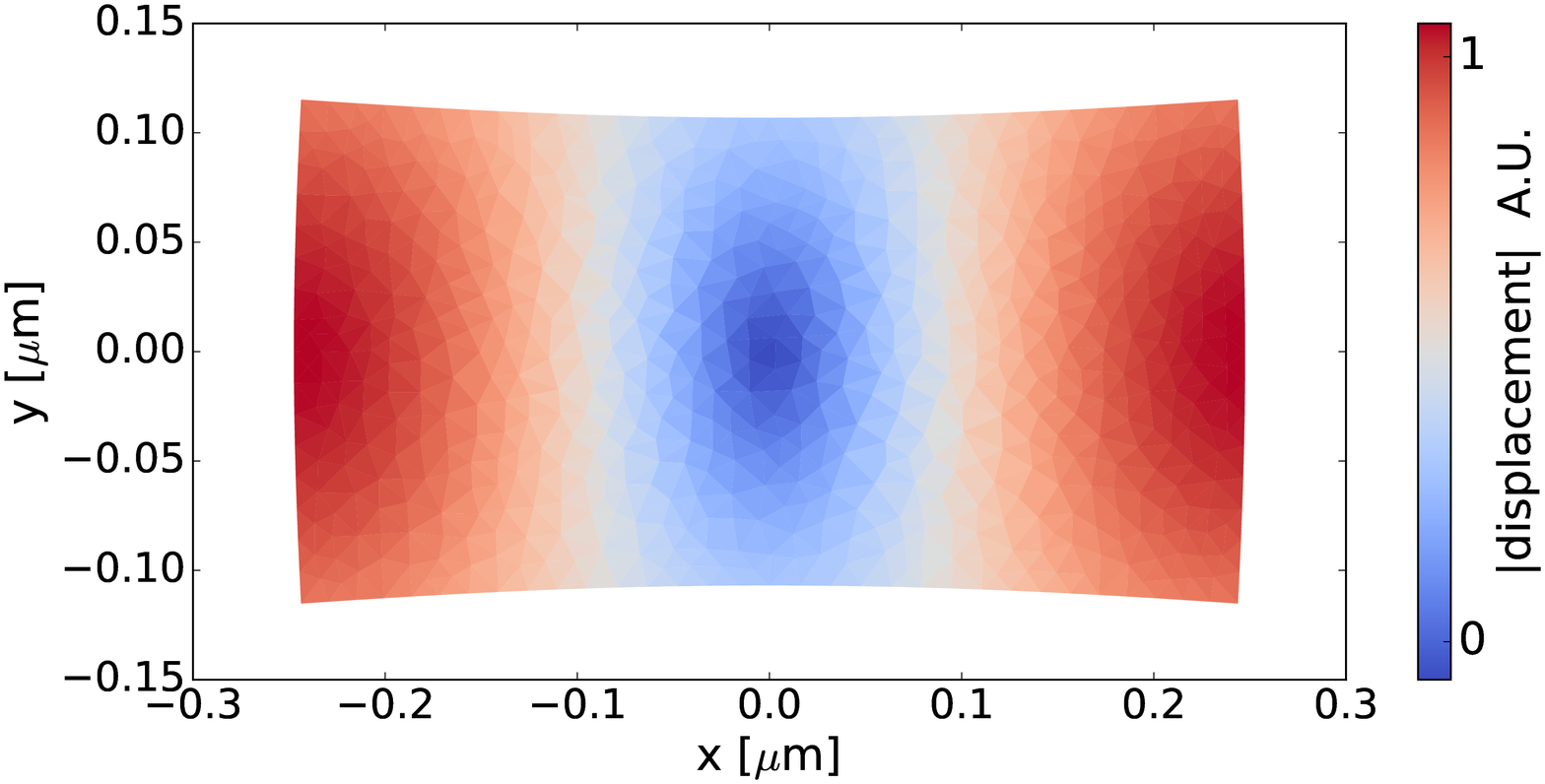}
   \caption{}
   \label{fig:mode1} 
\end{subfigure}
\vspace{-0.5cm}
\begin{subfigure}[b]{0.55\textwidth}
   \includegraphics[width=1\linewidth]{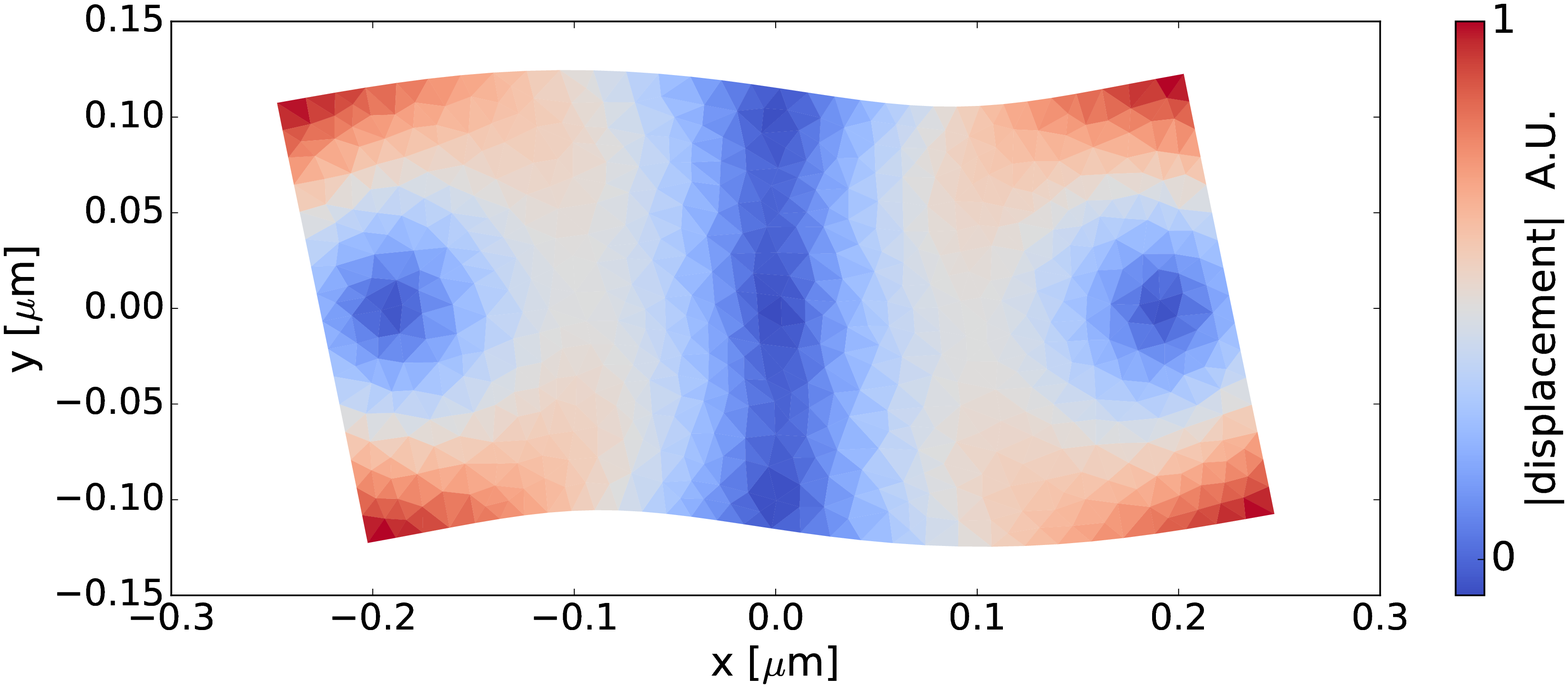}
   \caption{}
   \label{fig:mode2}
\end{subfigure}

\caption{ (a) The fundamental {\it x}-shear mode for an SOI waveguide in air; (b) The fundamental {\it x}-shear mode for an SOI waveguide in air.}
\end{figure}

The anisotropic nature of silicon necessitates the use of a fully-tensorial description of elasticity, as implemented in this work.  As mentioned before and following the MEMS literature \cite{hopcroft2010young}, some authors have chosen to use the simplified isotropic model for silicon \cite{rakich2012giant}. While it is qualitatively fair, the isotropic model can lead to large discrepancies in quantitative values. In extreme cases, when the waveguide is aligned along the [100] or the [110] crystal directions, the differences in eigenfrequencies is as much as 0.8 GHz, as shown in Table \ref{tab:silicon_eigenmodes}. The modal profile is included for completeness in Fig. \ref{fig:mode1}. Similar elastic anisotropy is expected from other cubic materials, such as germanium \cite{wolff2014germanium}.

\begin{table}[!t]
\renewcommand{\arraystretch}{1.3} 
\caption{Si Elastic Modes}
\label{tab:silicon_eigenmodes}
\centering
\begin{tabular}{|c|c|c|c| }
 \hline
  Mode & Fund. {\it x} shear & Fund. {\it x} shear & Torsional  \\
 \hline
  SBS & Forward & Forward & Forward  \\
  \hline
  Propagation axis & [100] & [110] & [100]  \\
  \hline
  Eigenvalue, this work & 8.4064 GHz & 9.2300 GHz &  10.633 GHz  \\
  \hline
  Eigenvalue, 3D COMSOL  & 8.4064 GHz & 9.2300  GHz &  10.633 GHz\\
  \hline
  Plot & Fig. \ref{fig:mode1} & - & Fig. \ref{fig:mode2}\\
  \hline 
\end{tabular}
\end{table}

\subsection{Brillouin Forces and Gain}
The presented elastic-wave mode-solver can be combined with an electromagnetic mode-solver, implemented in {\it FEniCS} and described in detail in \cite{logg2012automated}, to calculate the forces acting on the opttical waveguide. The electromagnetic mode-solver is based on minimizing the functional \cite{logg2012automated}
\begin{equation}
\begin{split}
F(E) = \int_{\Omega} \frac{1}{\mu_r}(\nabla_T \times E_T) \cdot (\nabla_T \times E_T) - k_0^2\epsilon_r E_T \cdot E_T + \\ \gamma^2[\frac{1}{\mu_r}(\nabla_T E_{z,\gamma} + E_T) \cdot (\nabla_T E_{z,\gamma} + E_T) -  k_0^2\epsilon_r E_z \cdot E_z] dx,
\end{split}
\label{eq:electromagnetic_functional}
\end{equation}
which comes from the wave equation with the $E = ( E_T + \gamma E_{z,\gamma})\exp(-\gamma z)$ ansatz. Since for propagating modes, $\gamma$ is imaginary, the transverse field is real and the {\it z}-field is imaginary. Also, $\epsilon_r $ and $\mu_r$ are assumed to be real and scalar.

There are three types of forces that contribute to the Brillouin gain \cite{qiu2013stimulated}. First, there is bulk electrostriction, which is the dominant force in optical fibers. Given that there are two fields present in  waveguides, described by $\boldsymbol{E} = \frac{1}{2}(\boldsymbol{E_p}\exp(ik_pz - \omega_pt) + \boldsymbol{E_s}\exp(ik_sz - \omega_st)) + c.c.$, the phase-matched component of the stress, $\sigma_{ij}^{ES}\exp(iq_bz -\omega_{mech} t)$, is given by
\begin{equation}
\sigma_{ij}^{ES} = -\frac{1}{4}\epsilon_0 n^4 p_{ijkl} (E_{pk}E_{sl}^* + E_{pl}E_{sk}^*),
\end{equation}
where $p$ is the electrostriction tensor and the force is 
\begin{equation}
 \boldsymbol{ f^{ES}} = -\nabla_T \cdot  \boldsymbol{\sigma^{ES}} - iq_b \boldsymbol{\hat{z}} \cdot  \boldsymbol{\sigma^{ES}}.
\label{eq:bulk electrostriction}
\end{equation}

Second, there is a boundary electrostriction force at the interface of dielectrics 1 and 2, given by 
\begin{equation}
{f_i^{BES}} = (\sigma_{1ij}^{ES} - \sigma_{2ij}^{ES}) n_j,
\end{equation}
where $n_j$ is a normal pointing from material 1 to material 2. The third force, arising from radiation pressure, also appears only on the boundary and can be computed from Maxwell's stress tensor. Again, retaining only the phase-matched components, $T_{ij}\exp(iq_bz -\omega_{mech} t)$, with
\begin{equation}
T_{ij} =  \frac{1}{2}\epsilon_0 \epsilon_r ((E_{pi}E_{sj}^* + E_{pj}E_{si}^*) - \delta_{ij} (E_{pk}E_{sk}^*)),
\end{equation}
the corresponding force from radiation pressure is given by
\begin{equation}
{f_i^{RP}} = (T_{2ij} - T_{2ij}) n_j.
\end{equation}

Distribution of the forces and the shape of the elastic modes is necessary for the computation of the Brillouin gain coefficient. Since Brillouin coupling coefficient can be related to the photon generation rate through particle flux conservation \cite{rakich2012giant} and  the phonon generation rate is proportional to the power generated by optical forces, $\int_{\Omega} \boldsymbol{f} \cdot \partial_t \boldsymbol{u^*} dx$, the gain coefficient is expressed as \cite{qiu2013stimulated}
\begin{equation}
G = \frac{\omega_{opt} Q_{mech}}{4 P_{mech} P_s P_p} |\int_{\Omega} \boldsymbol{f} \cdot  \boldsymbol{u^*} dx|^2, 
\label{eq:gain}
\end{equation}
where $\omega_{opt}$  and $\omega_{mech}$ refer to optical and mechanical frequencies of propagating waves, $Q_{mech}$ is the mechanical quality-factor of the mode, $P_{s,p} =  \frac{c_0}{2n_{g}}\epsilon_0 \int_{\Omega}  \boldsymbol{E_{s,p}} \cdot \boldsymbol{\epsilon_r E^*_{s,p}} dx$ are the optical powers of the pump scattered field and $P_{mech}  = \frac{1}{2}\omega_{mech}^2 \int_{\Omega} \rho \boldsymbol{u} \cdot \boldsymbol{u^*} dx$ is the mechanical power. The forces in the numerator are added coherently, i.e., $\boldsymbol{f} = \boldsymbol{f^{ES}}+\boldsymbol{f^{RP}}$.

\begin{figure}[!h]
\centering
   \includegraphics[width=1\linewidth]{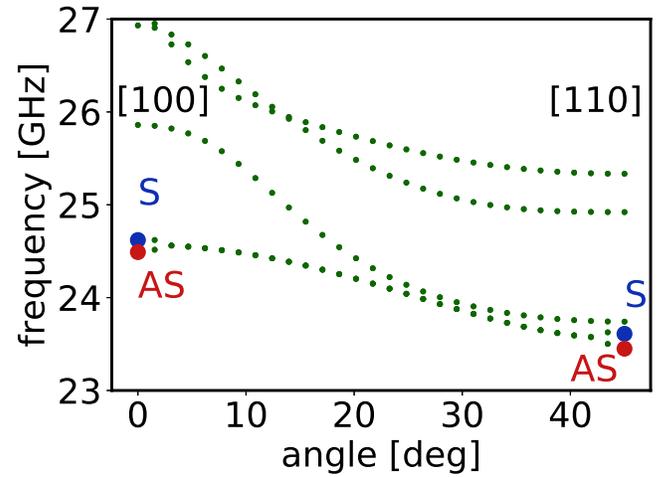}
   \caption{Resonant elastic frequencies of an silicon waveguide, suspended in air, for various crystal orientations. For an arbitrary direction, the waveguide is not symmetric elastically, thus merging of the symmetric (S) and asymmetric modes (AS) is observed.}
   \label{fig:dispersion curve} 
\end{figure}

\begin{figure*}

\begin{subfigure}[b]{0.33\textwidth}
\centering
   \includegraphics[width=1\linewidth]{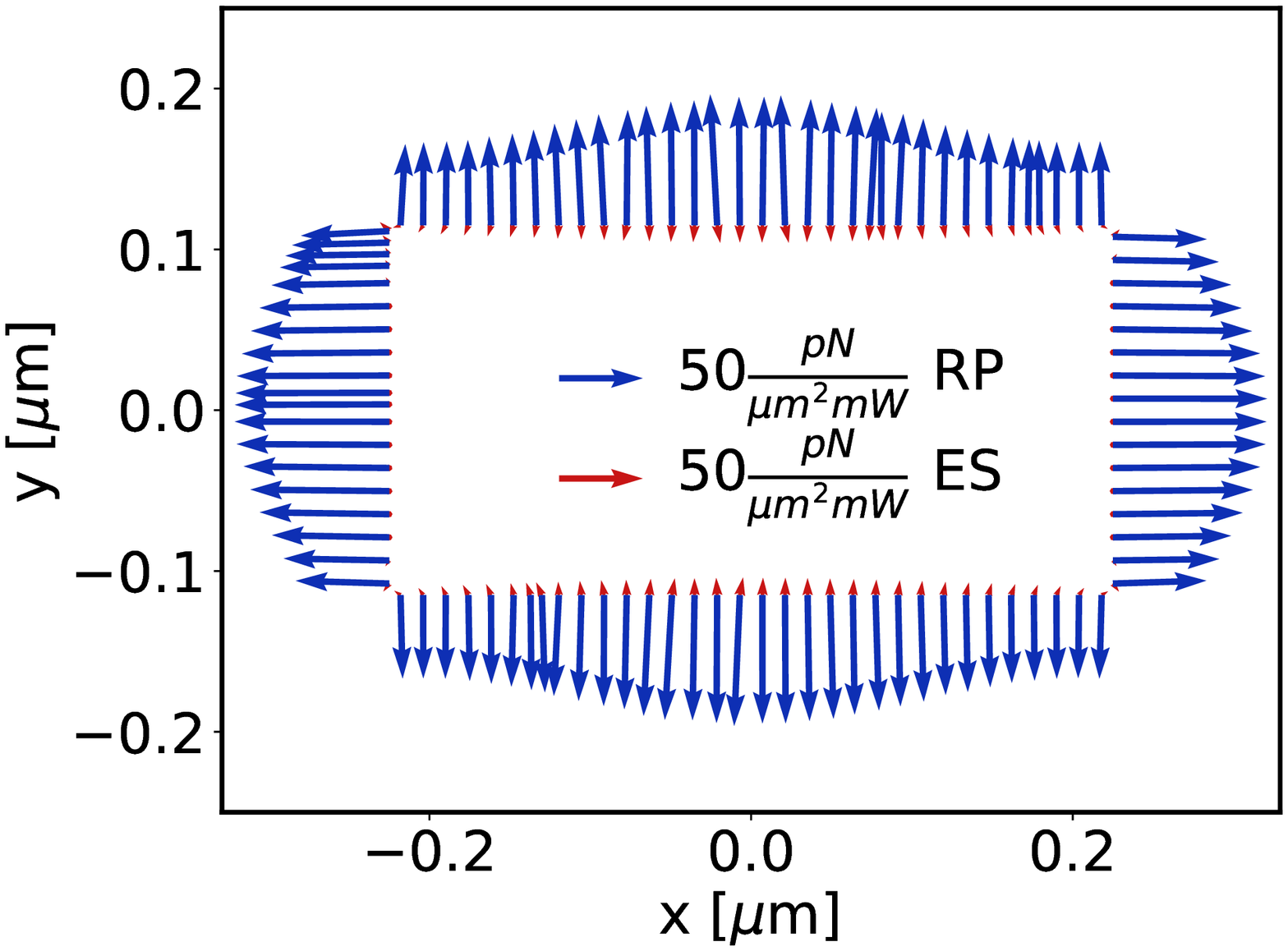}
   \caption{Forward SBS, boundary forces}
   \label{fig:bdr_fsbs} 
\end{subfigure}
\begin{subfigure}[b]{0.33\textwidth}
\centering
   \includegraphics[width=1\linewidth]{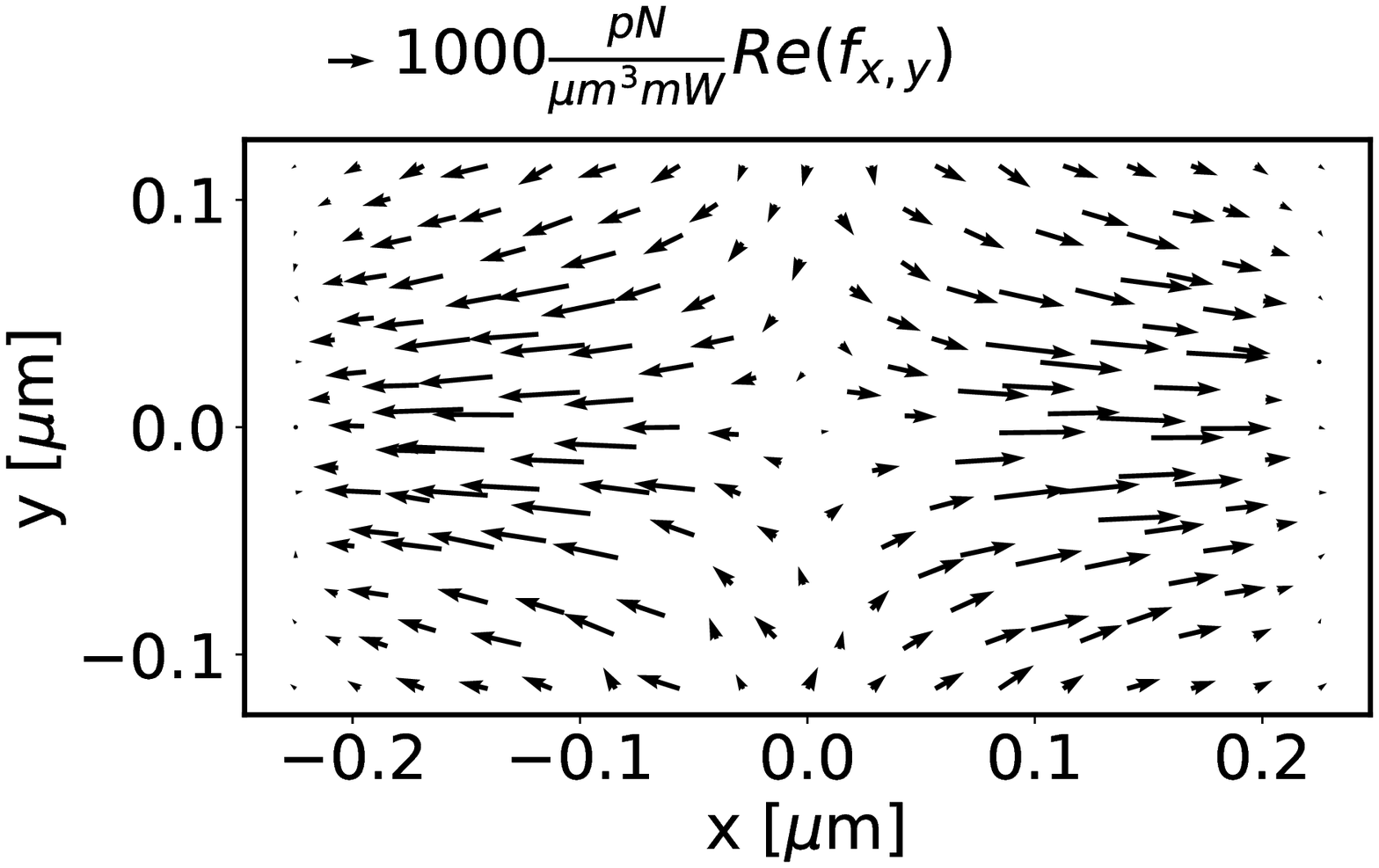}
   \caption{Forward SBS, bulk ES}
   \label{fig:bulk_fsbs} 
\end{subfigure}
\begin{subfigure}[b]{0.33\textwidth}
\centering
   \includegraphics[width=1\linewidth]{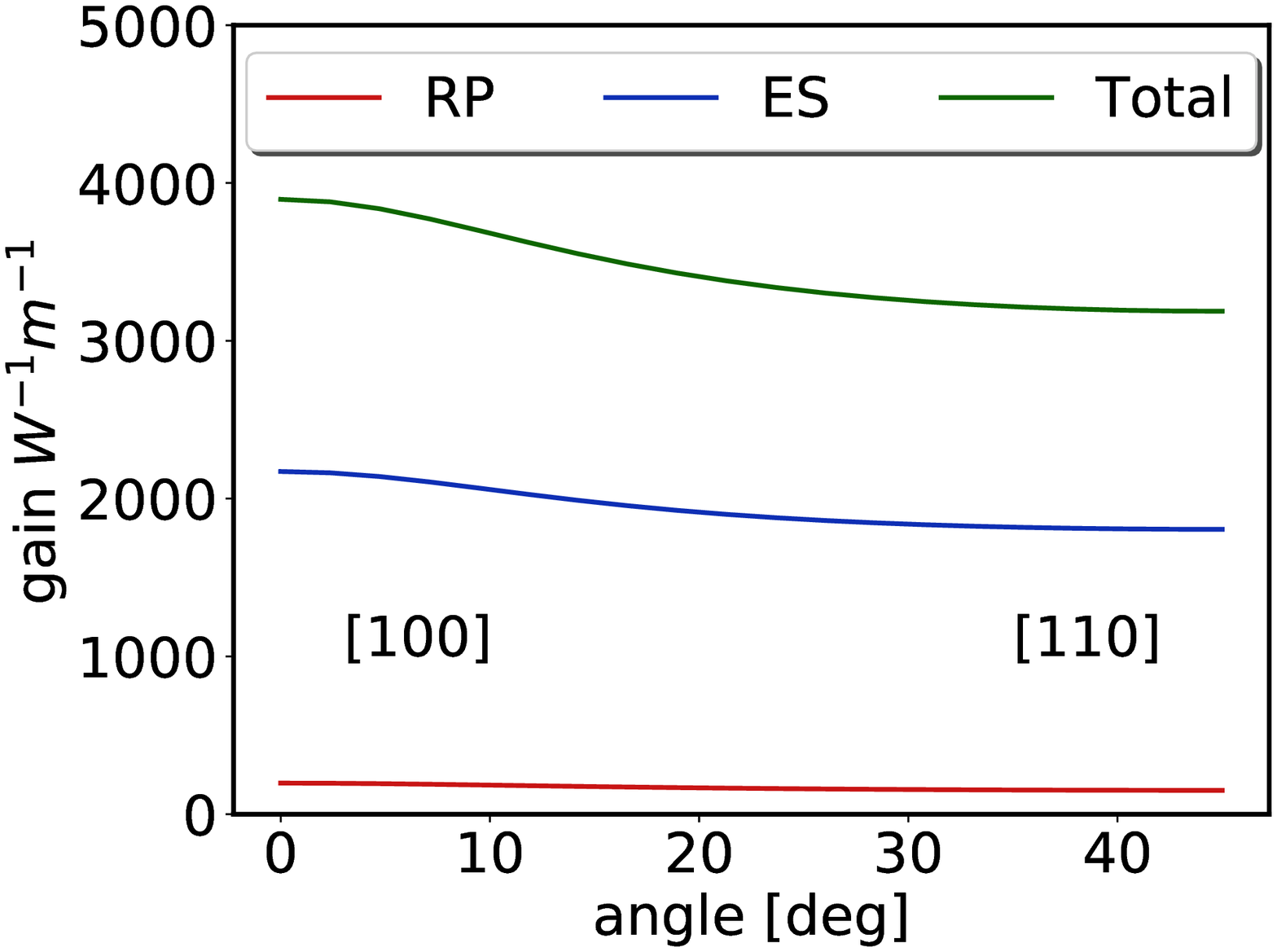}
   \caption{Forward SBS gain}
   \label{fig:fsbs_gain}
\end{subfigure}
\begin{subfigure}[b]{0.33\textwidth}
\centering
   \includegraphics[width=1\linewidth]{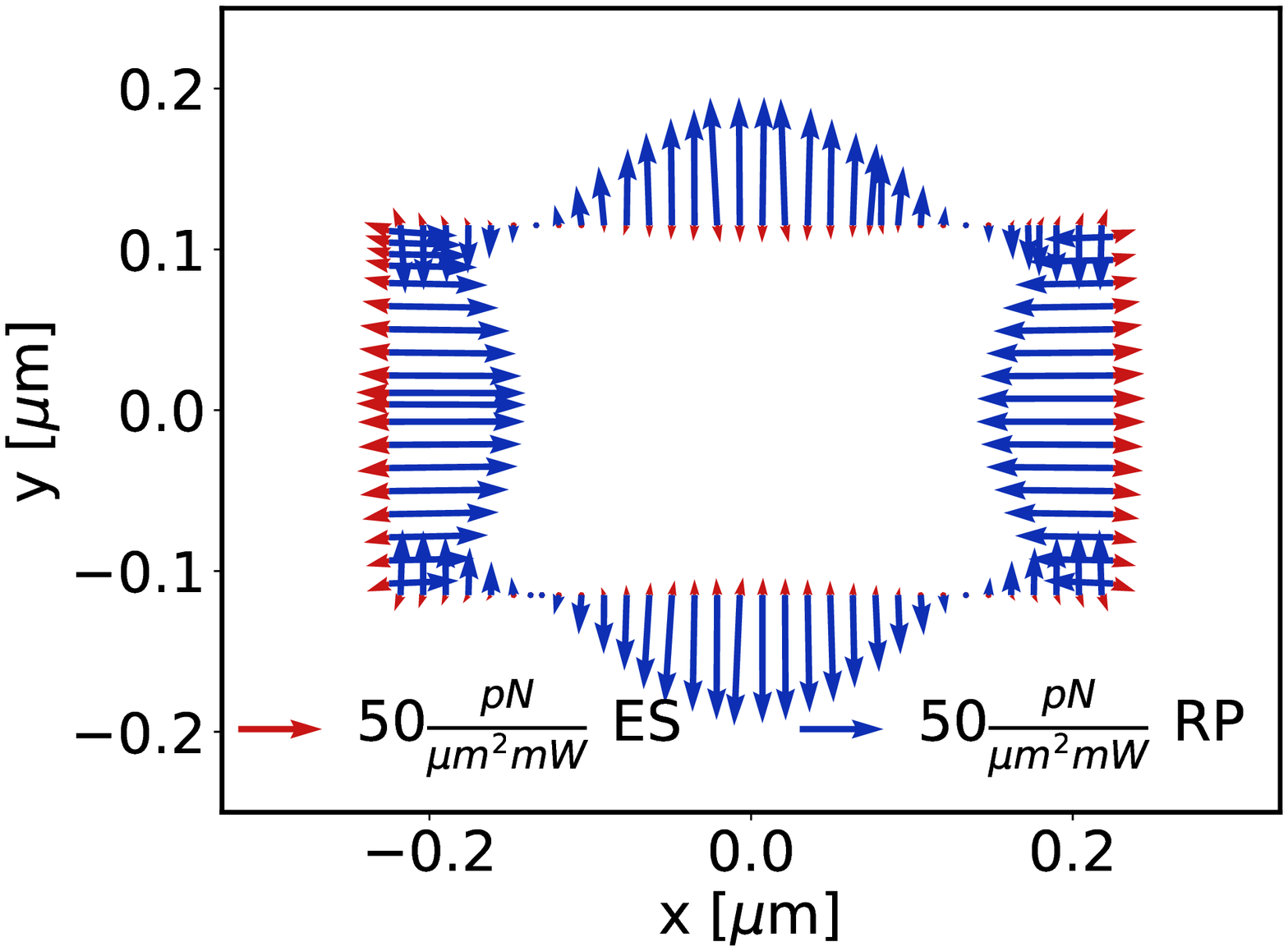}
   \caption{Backward SBS,  boundary forces}
   \label{fig:bdr_bsbs}
\end{subfigure}
\begin{subfigure}[b]{0.33\textwidth}
\centering
   \includegraphics[width=1\linewidth]{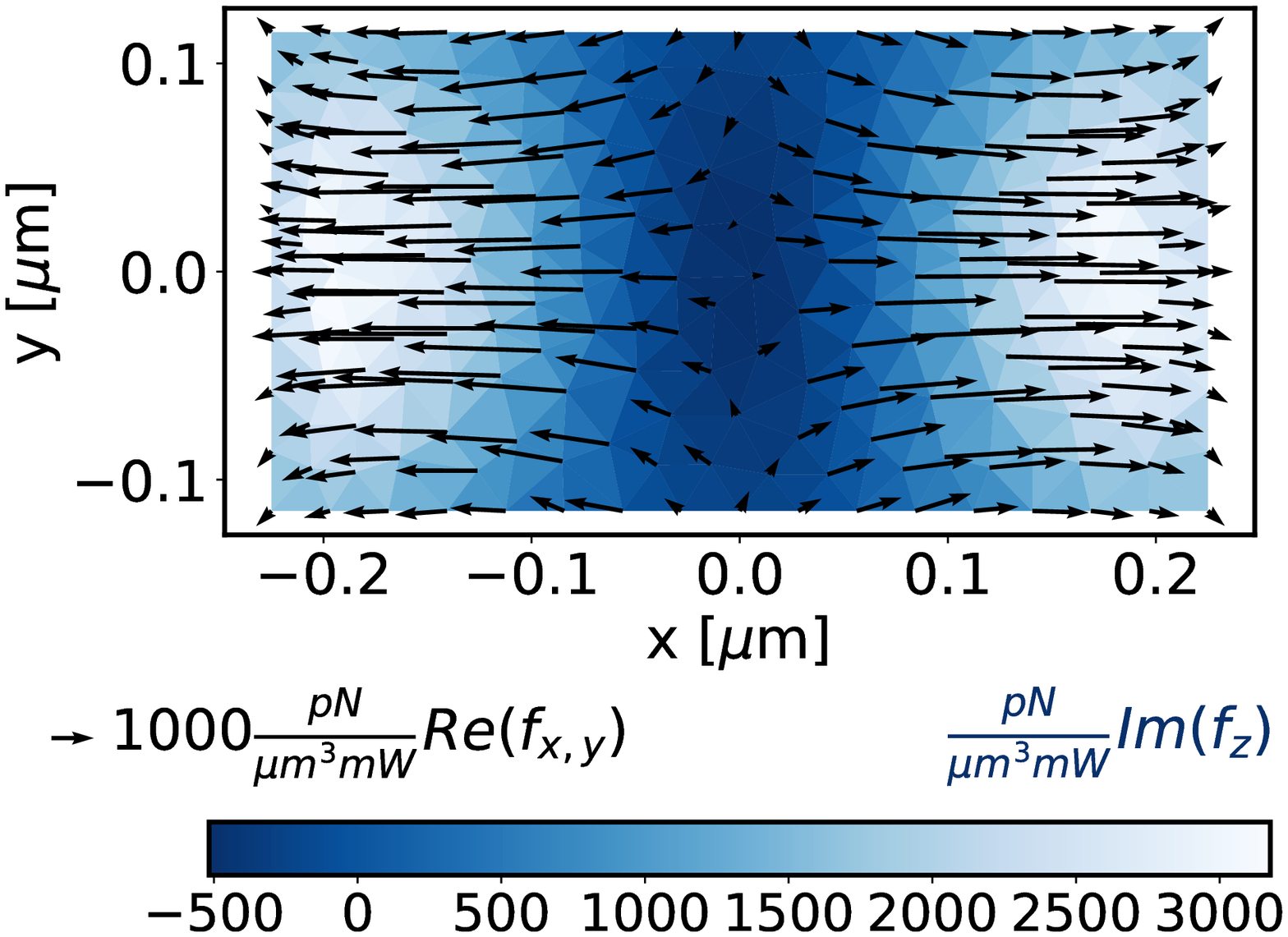}
   \caption{Backward SBS, bulk ES}
   \label{fig:bulk_bsbs}
\end{subfigure}
\begin{subfigure}[b]{0.33\textwidth}
\centering
   \includegraphics[width=1\linewidth]{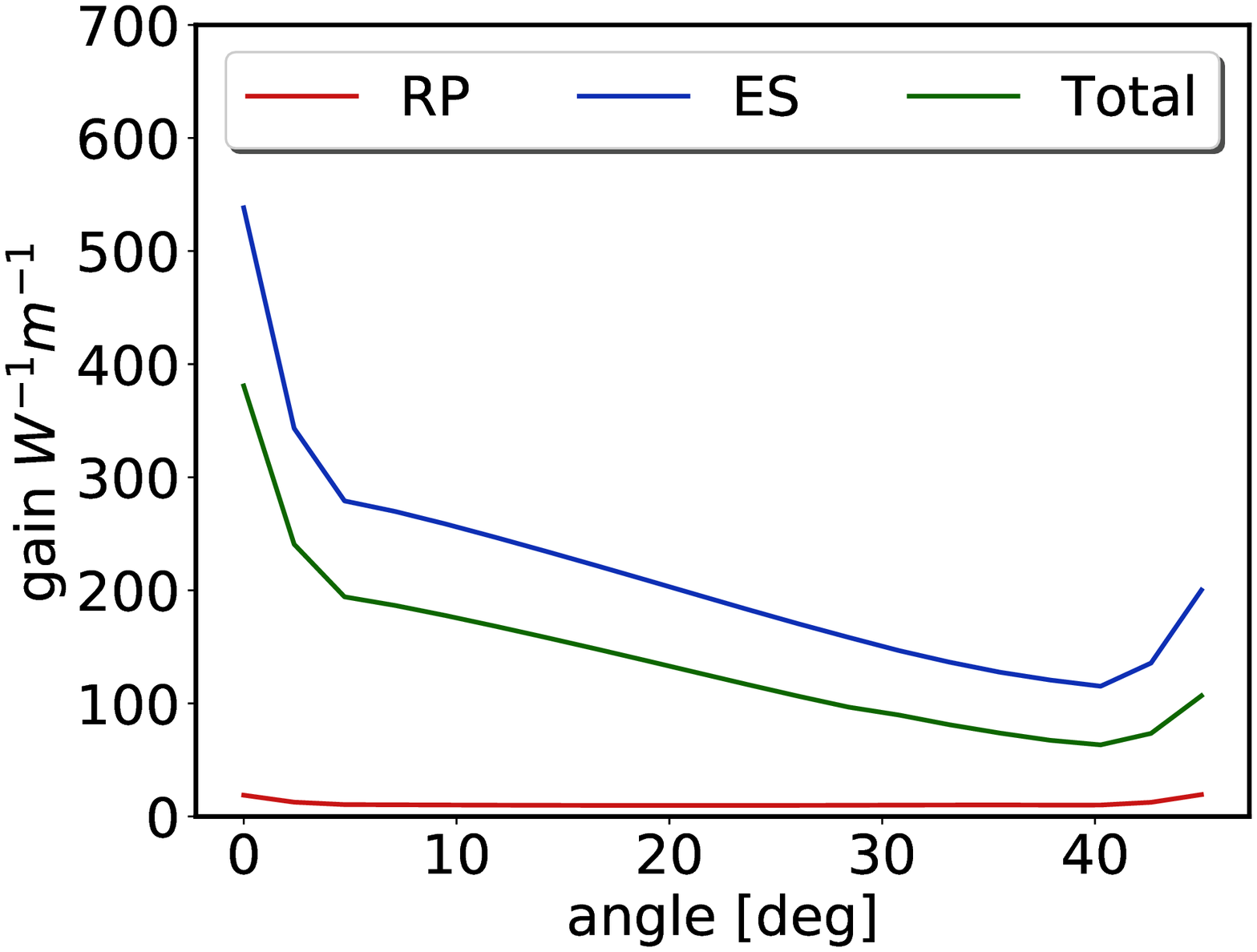}
   \caption{Backward SBS gain }
   \label{fig:bsbs_gain}
\end{subfigure}
\caption{Comparison of bulk and boundary forces and resulting gain for forward and backward Brillouin scattering originating from the fundamental TE mode and acting on a silicon waveguide suspended in air. The coupling is computed for the fundamental {\it x}-shear mode from Figure \ref{fig:mode1} and different crystal orientations. Bulk electrostriction forces are in plotted in black, the  boundary radiation pressure in blue and boundary electrostriction in red. In the case of back scattering in Figure (e) bulk electrostriction has a large imaginary component acting in the z direction in addition to lateral components. The gain dependence on crystal orientation shows discontinuities due to mode merging as explained in detail in text.}
\label{fig:forces}
\end{figure*}

The simulations in Fig. \ref{fig:forces} are performed for a rectangular silicon waveguide (450 nm $\times$ 230 nm), suspended in air, and for the fundamental TE mode at the 1550 nm wavelength. We have assumed the photoelastic constants of p$_{11}$ = -0.09, p$_{12}$ = 0.017,  p$_{44}$ = -0.051 and $Q_{mech}$ = 249  \cite{van2015interaction}.  It is important to distinguish two cases. For the forward Brillouin scattering, $\boldsymbol{E_s} = \boldsymbol{E_p}$ and $q_b \approx 0$, since the frequency shift is small in comparison to the optical carrier frequency. In this case (Figs. \ref{fig:bdr_fsbs} and  \ref{fig:bulk_fsbs}), all forces are real.  The boundary electrostriction (ES) is weaker (11\% when comparing the maxima) and in opposite direction to the radiation pressure (RP).
For the [110] direction of propagation the dominant {\it x}-components of bulk electrostriction (1684 W$^{-1}$m$^{-1}$) and radiation pressure  on the boundary (151 W$^{-1}$m$^{-1}$) add up constructively resulting in a total gain of 3099 W$^{-1}$m$^{-1}$. This value is in good agreement with the experimental direct characterization of the gain yielding 3218 W$^{-1}$m$^{-1}$ and indirect characterization through cross-phase modulation giving 3055 W$^{-1}$m$^{-1}$ \cite{van2015interaction}. The total simulated gain for the [100] direction is  3896  W$^{-1}$m$^{-1}$. This is to be expected from the $1/\omega_{mech}^2$ dependence of gain in Equation \ref{eq:gain} and the aforementioned difference of 0.8 GHz in resonance frequency between the two directions of propagation.  Evidently, using the described fully-tensorial model is critical in modeling SBS in devices constructed from anisotropic materials. 

For the backward Brillouin scattering, $\boldsymbol{E_s} = \boldsymbol{E_p^*}$ and $q_b = 2k_0n_{eff}$. In the software, the employed sign convention is negative wavevector, $-\gamma$, for the pump and positive, $+\gamma$, for the scattered field, which also implies positive wavevector, $+q_b$, for the elastic wave. In Fig. \ref{fig:bdr_bsbs}, the boundary electrostriction (ES) is weaker (29\% when comparing the maxima) and in opposite direction to the radiation pressure (RP). Since the dominant {\it x}-component of radiation pressure in Fig. \ref{fig:bdr_bsbs} acts in antiphase to the bulk electrostriction in Fig. \ref{fig:bulk_bsbs}, the total Brillouin gain is smaller than the electrostriction component alone, as shown in Fig. \ref{fig:bsbs_gain}. The value of total gain for the [100] direction is 380 W$^{-1}$m$^{-1}$. 

The features of backward SBS is even more complex than forward SBS and is best explained with the aid of a dispersion diagram from Fig. \ref{fig:dispersion curve}, where the  resonant frequencies are plotted versus the crystal orientation with respect to the direction of propagation. For the gain coupling calculation, we follow the fundamental symmetric mode marked with a blue "S" in Fig. \ref{fig:dispersion curve} at 24.513 GHz and $Q_{mech}$ = 249. We note that, in general, the elastic mode is not symmetric for cubic materials such as silicon \cite{su2017theoretical}, unlike the optical mode. Thus, the initially symmetric mode acquires a degree of asymmetry as the crystal orientation is tilted, to the point that at 3$^o$, the symmetric- and the antisymmetric-mode branches merge. This is the origin of the first discontinuity in the gain diagram in Fig. \ref{fig:bsbs_gain}. The next discontinuity at 42$^o$ occurs when the common branch splits again into the symmetric and antisymmetric modes. The optical mode is symmetric, thereby the optical force distribution is symmetric, so the coupling to the asymmetric modes is null. Therefore, the  Brillouin gain drops as the crystal orientation is rotated away from the [100] or [110] directions, since the elastic modes acquire a degree of asymmetry.

\section{Conclusions}
In summary, we have presented a fully-tensorial elastic-wave mode-solver and discussed its implementation in the open-source finite element solver, \textit{FEniCS}. The source code is available at \url{https://github.com/MarcinJM/PySBS}. The use of 2D SAFE method in this work leads to computational times smaller by two orders of magnitude than the commercial 3D finite-element solver, COMSOL$^{TM}$. We have also performed a thorough review and comparison of a multitude of approximations used in simulating the elastic modes in the SBS literature and compared them against the present fully-tensorial model. In some cases, the discrepancy in the calculated eigenvalues was found to be as large as 0.8 GHz. We have shown that for silicon with arbitrary crystal orientations, the elastic modes do not need to have the same symmetry as the optical modes, which greatly affects the Brillouin coupling coefficient. The modeling tool is made publicly available to other researchers and is expected to be useful in understanding and tailoring SBS in integrated devices.

\ifCLASSOPTIONcaptionsoff
  \newpage
\fi

\bibliographystyle{IEEEtran}
\bibliography{references}{}

\end{document}